# Simple derivation of Schrödinger equation from Newtonian dynamics

Michele Marrocco

*Dipartimento di Fisica, Università di Roma 'La Sapienza'*

*P.le Aldo Moro 5, I-00185 Rome, Italy*

*&*

*ENEA*

*(Italian National Agency for New Technologies, Energies and Sustainable Economic*

*Development)*

*via Anguillarese 301, I-00123 Rome, Italy*

## Abstract

The Eherenfest theorem states that Schrödinger representation of quantum mechanics (wave mechanics) reproduces Newton's laws of motion in terms of expectation values. Remarkably, the contrary is considered elusive and, indeed, many authors have tried to obtain wave mechanics starting from other alternative frameworks of classical mechanics (for instance, Hamilton-Jacobi theory). Despite this common opinion, we present here a simple method to make Newtonian dynamics develop naturally into Schrödinger representation. The proof is based on the assumption of matter waves and is laid out in three fundamental steps. First, the role of classical density functions is underlined in view of their use to define constants of the motion for massive particles. Thanks to this preparatory step, density functions generate wave-functions whose spatial and time variables obey Newton's laws of motion. The resulting wave equation is defined in dependence on a parameter that plays the identical role of the constant $K$ introduced by Schrödinger in the original formulation of his theory. In the final step, the classical wave equation is treated under the hypothesis of conservative forces common to the Eherenfest theorem and, after some algebra, the Schrödinger equation emerges by means of the identification of the classical momentum with de Broglie momentum of matter waves.

## I. INTRODUCTION

One of the main routes to quantum mechanics runs through the Schrödinger equation and the concept of wave function.[1] Despite the undoubted importance of this cornerstone of modern physics, the subject is ordinarily introduced in courses on quantum mechanics without much detail about its conceptual foundations, with the result that, according to prominent physicists,[2] the derivation of the Schrödinger equation conveys a sense of dissatisfaction. In reality, the disappointment has something to do with the heuristic explanation reported by Schrödinger in his original conjecture.[3] Its weakness is, for instance, underlined by Feynman in the third volume of the Lectures where one can read the following comment about the founder of wave mechanics: "some of the arguments he used were even false, but that does not matter; the only important thing is that the ultimate equation gives a correct description of nature".[4] Against this compliant attitude, Feynman himself provided us with a derivation of the Schrödinger equation that led him to the path integral formulation of quantum field theory.[5, 6] Responding to the same stimulus and in the hope of a sound explanation of the equation, many other physicists have come up with several proposals that span the last 50 years.[2, 7-23] Along this line of research, the current attempt aims at the definition of a simple and, at the same time, rigorous approach to the task. But, before going into the details and to better contextualize our purpose, let us make some preliminary remarks concerning ordinary approaches to the Schrödinger picture of quantum mechanics.

In general, the introduction to this fundamental subject of modern physics is made on the basis of three mutually exclusive didactic criteria. The simplest responds to the legitimate claim that the Schrödinger equation is so well known that its detailed derivation can be neglected in favor of its solutions to fundamental problems.[24, 25] Under this orientation, the equation is simply assumed as a fact. On the other hand, the undisputable demand for a deeper physical insight urges us to provide a theoretical minimum at least.[26] To this end, the most common approach draws its inspiration from plausibility arguments[27-29] that are simpler analogs of the original idea developed by Schrödinger. Beyond their undeniable strength of pedagogical interest, such proposals are not without shortcomings (e.g., restriction to the free particle, assumption of a constant potential) and, for this reason, they are not good enough for whoever is in pursuit of a flawless method to secure the Schrödinger equation against any doubt about its theoretical foundation. From this perspective, a considerable number of authors has met the challenge of an exact formulation from first principles.[2, 7-23] Among these attempts, the Hamilton-Jacobi theory of classical mechanics has attracted a lot of attention because of the guidance it gave to Schrödinger in the discovery of his equation. Unlike this acknowledged theoretical standpoint, which presents itself with nonlinearities that must be suppressed in some manner (see, for instance, the discussion in Ref. 2), here we take the alternative view, suggested by some authors,[7, 13] that the Schrödinger equation is directly achievable through Newton's laws of dynamics.

The connection was clear soon after the birth of modern quantum mechanics (Ehrenfest theorem)[30] and, nowadays, it is common knowledge that the Schrödinger equation "plays a role logically analogous to Newton's second law".[25] Thus, the objective of this work is to prove that the correspondence is stronger than an analogy. However, it must be remarked that the current proof has the fundamental premise in the de Broglie postulate of matter waves. In this regard, our starting point coincides with the assumption from which Schrödinger moved in search of the equation that describes waves peculiar to massive particles.[3] In addition, the postulate has the further benefit of avoiding the stochastic forces that characterize former proposals made to show that the radical departure from classical physics is unnecessary.[7, 13]

The work is organized as follows. In Section II, density functions are introduced in view of their classical description of constants of motion of massive particles. It is then shown that such functions can be rewritten in terms of eigenfunctions that provide solutions to the eigenvalue problem established for a proper linear operator. In Section III, the spatial and time dependences appearing in the eigenfunctions are treated according to the second law of Newtonian dynamics and, in this manner, a classical wave equation is generated. In Section IV, considering the reference to conservative forces that are central to the Ehrenfest theorem,[30] the classical wave equation is combined with the divergence theorem of three-dimensional calculus and, in the end, the Schrödinger equation is recovered under the assumption of the de Broglie postulate. The final Section V is relative to the conclusions.

## II. CONSTANTS OF THE MOTION AND EQUIVALENCE BETWEEN CLASSICAL DENSITY FUNCTIONS AND LINEAR OPERATORS

The approach considered here begins with the introduction of a three-dimensional Cartesian reference frame within which a classical non-relativistic particle of mass $m$ and momentum $\mathbf{p}$ follows a certain trajectory that is solution to the equation $m\ddot{\mathbf{r}} = \mathbf{F}$ (the dots indicate time derivatives while $\mathbf{F}$ is the force acting at the spatial point determined by the vector $\mathbf{r}$). The space of volume $V$ is supposed continuous and, for example, we can identify a suitable density function $\rho_C(\mathbf{r},\mathbf{p},t)$ whose spatial integration gives rise to a constant of the motion so that

$$C = \int_V \rho_C(\mathbf{r},\mathbf{p},t) d\mathbf{r} \tag{1}$$

where all the dynamical dependences have been considered in the argument of $\rho_C(\mathbf{r},\mathbf{p},t)$. Typical examples of density functions that satisfy Eq. (1) are the mass density and the energy density of a particle. One can think of many other specific examples, but more generally we can ask ourselves what is entailed in Eq. (1) if we suppose that the same constant of motion has to characterize the quantum description. In trying to answer, we first rewrite Eq. (1) in another version more suitable for whoever is familiar to quantum mechanics.

Classically, the density function $\rho_C(\mathbf{r},\mathbf{p},t)$ can always be rendered as follows

$$\rho_C(\mathbf{r},\mathbf{p},t) = \gamma |\Gamma(\mathbf{r},\mathbf{p},t)|^2 \qquad (2)$$

where $\gamma$ is a number and $\Gamma(\mathbf{r},\mathbf{p},t)$ is a complex-valued function that contains the relevant dependences on the dynamical variables. We also suppose that $\Gamma(\mathbf{r},\mathbf{p},t)$ is normalized according to

$$\int_V |\Gamma(\mathbf{r},\mathbf{p},t)|^2 d\mathbf{r} = 1 \qquad (3)$$

which implies that $C = \gamma$. It is then easy to show that if we were able to define a linear operator $\hat{O}$ such that $\hat{O}\,\Gamma(\mathbf{r},\mathbf{p},t) = \gamma\,\Gamma(\mathbf{r},\mathbf{p},t)$ we would find that Eq. (1) becomes

$$C = \int_V \Gamma^*(\mathbf{r},\mathbf{p},t) \hat{O}\,\Gamma(\mathbf{r},\mathbf{p},t) d\mathbf{r} = \gamma \qquad (4)$$

where the superscript * indicates the complex conjugate. In conclusion, the classical description of Eqs. (1) and (2) is equivalent to the description in terms of linear operators that, in turn, form the reference mathematical tool of quantum mechanics. Vice versa, a dynamical description based on linear operators is closely connected to some density functions that are associated with some constants of the classical motion.

The equivalence summarized above imposes an important condition on the eigenfunction $\Gamma(\mathbf{r},\mathbf{p},t)$. To determine it, we first observe that, if the momentum changes with time, Eqs. (1) and (4) imply that the dependences on instantaneous momentum $\mathbf{p}$ and time $t$ must disappear if the motion has to be defined by its constant $C$. This means that for a given momentum $\mathbf{p}_0$ belonging to the range of possible values of $\mathbf{p}$, it is always possible to find one or more points $\mathbf{r}_j = \mathbf{r}(t_j)$ (with $j = 1, 2,...$ a positive index) that satisfy the following equation

$$C = \gamma V \left|\Gamma(\mathbf{r}_j,\mathbf{p}_0,t_j)\right|^2 \qquad j = 1, 2,... \tag{5}$$

which results from the mean value theorem for the definite integrals of Eq. (3). Given the equality $C = \gamma$ obtained before, Eq. (5) leads to the condition

$$V \left|\Gamma(\mathbf{r}_j,\mathbf{p}_0,t_j)\right|^2 = 1 \tag{6}$$

that introduces a constraint on the shape of the eigenfunction $\Gamma(\mathbf{r},\mathbf{p},t)$. The constraint has a simple representation when we solve Eq. (6) as follows

$$\Gamma(\mathbf{r}_j,\mathbf{p}_0,t_j) = \frac{1}{V^{1/2}} e^{i\alpha \mathbf{r}_j \cdot \mathbf{p}_0} \tag{7}$$

where $\alpha$ is a proportionality constant with physical units of the inverse of an action. Eq. (7), although limited by the condition $\mathbf{r}(t_j) = \mathbf{r}_j$, can be used to determine the more general shape of $\Gamma(\mathbf{r},\mathbf{p},t)$. A very simple application is when the momentum remains constant at any time (i.e., free particle) so that we can replace the discrete values of $\mathbf{r}_j$ and $t_j$ with their continuous representations. Then, based on Eq. (7), the eigenfunction $\Gamma(\mathbf{r},\mathbf{p},t)$ is identified by the plane wave when the momentum takes the only available value of $\mathbf{p}_0$

$$\Gamma(\mathbf{r},\mathbf{p}_0,t) = \frac{1}{V^{1/2}} e^{i\alpha \mathbf{r}\cdot\mathbf{p}_0}. \tag{8}$$

Understandably, the periodic condition

$$e^{i\alpha (\mathbf{r}+\Delta\mathbf{r})\cdot\mathbf{p}_0} = e^{i\alpha \mathbf{r}\cdot\mathbf{p}_0} \tag{9}$$

where $\alpha \Delta\mathbf{r} \cdot \mathbf{p}_0 = 2\pi$ (with $\Delta\mathbf{r}$ a spatial vector in the direction of the motion) results in the definition of the wavelength $\lambda = |\Delta\mathbf{r}|$

$$\lambda = \frac{2\pi}{\alpha\, p_0} \tag{10}$$

where $p_0 = |\mathbf{p}_0|$ and, as a direct consequence, we can easily remark that $\lambda$ becomes the de Broglie wavelength $\lambda_{DB}$ of the free particle as soon as we set $\alpha = 1/\hbar$.

A further and well-known consequence of the constant value of $\mathbf{p}_0$ appears when we interpret Eq. (1) as one component of a conserved vector and, for the case under consideration where $\mathbf{C} = (C_x, C_y, C_z) = \mathbf{p}_0$, then Eq. (4) reads

$$\mathbf{p}_0 = \int_V \Gamma^*(\mathbf{r},\mathbf{p},t)\hat{\mathbf{P}}\, \Gamma(\mathbf{r},\mathbf{p},t) d\mathbf{r} = \frac{1}{V}\int_V e^{-i\alpha \mathbf{r}\cdot\mathbf{p}_0}\hat{\mathbf{P}}\, e^{i\alpha \mathbf{r}\cdot\mathbf{p}_0} d\mathbf{r} \tag{11}$$

that is solved if we take

$$\hat{\mathbf{P}} = -\frac{i}{\alpha}\nabla. \qquad (12)$$

Eq. (12) shows the usual quantum-mechanical representation of the momentum operator once we let $\alpha = 1/\hbar$ as suggested before.

Armed with these very elementary findings (i.e., wave function of a free particle and representation of the momentum operator) that are however part of the initial concepts offered in courses on quantum mechanics, the next level of our itinerary is one step closer to the point we are trying to make here.

## III. DYNAMICAL LAW OF A NEWTONIAN PARTICLE WITHOUT ACCURATE KNOWLEDGE OF ITS MOMENTUM AND POSITION

The correspondence between density functions and linear operators, briefly outlined before, leads us to some basic and very simple notions of Schrödinger theory for a free particle as soon as we accept the hypothesis of matter waves described by the de Broglie postulate. However, despite the general validity of Eqs. (6) and (7) for a chosen value $\mathbf{p}_0$ within a set of possible values of the instantaneous momentum $\mathbf{p}$, we are not yet ready to work out the dependences appearing in $\Gamma(\mathbf{r},\mathbf{p}_0,t)$ when the particle interacts with a potential $U$ associated with the force $\mathbf{F}$. Indeed, the plane wave of Eq. (8) applies only if the momentum is constant. Instead, if the momentum is time dependent, the continuous limit of Eq. (7) cannot be taken and more general functions $\Gamma(\mathbf{r},\mathbf{p}_0,t)$ have to be found. To this end, we look for solutions of the kind

$$\Gamma(\mathbf{r},\mathbf{p}_0,t) = \frac{1}{V^{1/2}} f(\xi) = \frac{1}{V^{1/2}} f(\alpha\,\mathbf{r}\cdot\mathbf{p}_0) \tag{13}$$

where the definition of $\xi = \alpha\,\mathbf{r}\cdot\mathbf{p}_0$ is suggested by the conclusions reached for the function $\Gamma(\mathbf{r},\mathbf{p}_0,t)$ of the free particle seen in Eq. (8) so that, when $\mathbf{r}(t_j) = \mathbf{r}_j$, Eq. (6) becomes

$$\left|f(\xi_j)\right|^2 = \left|f(\alpha\,\mathbf{r}_j\cdot\mathbf{p}_0)\right|^2 = 1 \tag{14}$$

which agrees with Eq. (7) if we set $f(\xi_j) = f(\alpha\,\mathbf{r}_j\cdot\mathbf{p}_0) = Exp(i\alpha\,\mathbf{r}_j\cdot\mathbf{p}_0)$. Given this fundamental condition, the further step is the characterization of $f(\xi)$ that, apart from Eq. (14), remains not well defined according to the information we gained until now.

First of all, we observe that the variable $\xi$ is convenient because, as soon as we consider the evolution of $f(\xi)$, its time derivative depends on the time derivative of $\xi$ which, in turn, is connected to the kinetic energy $|\mathbf{p}_0|^2 /(2m)$ when $\dot{\mathbf{r}} = \mathbf{v}_0 = \mathbf{p}_0 / m$. Shortly, we are going to see how this information can be turned to our advantage. In addition, the role played here by the constant $\alpha$ is identical to the role given by Schrödinger to the constant $K$ appearing in the Hamilton principal function $S = K\,log\,\psi$ with $\psi$ the later known Schrödinger wave function [see Eq. (2) in the first paper of Ref. 3] and, in complete analogy with the use of $K$ made by the founder of wave mechanics, we will treat our constant $\alpha$ as a free parameter whose value will be determined by the evidence of matter waves with momenta given by the de Broglie equation.

Although the function $f(\xi)$ has been introduced without any specific definition of its properties (except for Eq. (14) and the above-mentioned definition of $\xi$), some conditions restrict the choice. First of all, its evolution depends on the vector $\mathbf{r}(t)$ extracted from the second Newton's law $m\ddot{\mathbf{r}} = \mathbf{F}$. To determine this evolution, we can calculate the first two time derivatives of $f(\xi)$

$$\frac{df}{dt} = \alpha \frac{df}{d\xi} \frac{d}{dt}(\mathbf{p}_0 \cdot \mathbf{r}) = \alpha \mathbf{p}_0 \cdot \dot{\mathbf{r}} \frac{df}{d\xi} \qquad (15)$$

$$\frac{d^2 f}{dt^2} = \alpha^2 (\mathbf{p}_0 \cdot \dot{\mathbf{r}})^2 \frac{d^2 f}{d\xi^2} + \alpha \mathbf{p}_0 \cdot \ddot{\mathbf{r}} \frac{df}{d\xi} \qquad (16)$$

and considering that

$$\nabla f = \alpha \, \mathbf{p}_0 \frac{df}{d\xi} \qquad (17)$$

$$\nabla^2 f = \alpha^2 \, p_0^2 \frac{d^2 f}{d\xi^2} \qquad (18)$$

with $p_0^2 = |\mathbf{p}_0|^2 = p_{0x}^2 + p_{0y}^2 + p_{0z}^2$, we find

$$-\frac{(\mathbf{p}_0 \cdot \dot{\mathbf{r}})^2}{p_0^2} \nabla^2 f + \frac{d^2 f}{dt^2} = \alpha \mathbf{p}_0 \cdot \ddot{\mathbf{r}} \frac{df}{d\xi} \qquad (19)$$

which establishes the dynamical law followed by the function $f(\xi)$ in order to carry the information about the state of the particle. The significance of Eq. (19) stems from the left-hand side where we have the description of a wave with an associated propagation speed of $(\mathbf{p}_0 \cdot \dot{\mathbf{r}})^2 / p_0^2$. The right-hand side can be regarded as the source term of the wave. Regardless of these details, we recover the known homogenous wave equation for free particles being $\ddot{\mathbf{r}} = 0$ and hence the right-hand side of Eq. (19) vanishing, that is

$$-\frac{(\mathbf{p}_0 \cdot \dot{\mathbf{r}})^2}{p_0^2} \nabla^2 f + \frac{d^2 f}{dt^2} = 0. \tag{20}$$

In view of its importance, this case is treated in the remaining part of this Section and Eq. (20) reduces to

$$\nabla^2 f - \frac{1}{|\mathbf{v}_0|^2} \frac{d^2 f}{dt^2} = 0 \tag{21}$$

where we have used the fact that the momentum the free particle had at the beginning is conserved and $\dot{\mathbf{r}} = \mathbf{p}_0 / m = \mathbf{v}_0$ at any time. Eq. (21) differs from the familiar wave equation by the total time derivative that replaces the more usual partial derivative. Although the total time derivatives appears in classical mechanics of waves in continuous media,[31] this little mismatch is removed by considering that $\xi = \alpha \mathbf{p}_0 \cdot \mathbf{r} = \alpha \mathbf{p}_0 \cdot \dot{\mathbf{r}} t = \alpha \mathbf{p}_0 \cdot \mathbf{v}_0 t = \Omega t$, with $\Omega = \alpha p_0^2 / m$ a constant angular frequency. Thus, the identity

$$\frac{df}{dt} = \frac{\partial f}{\partial t} \tag{22}$$

is trivially satisfied for the free particle. Furthermore, the time derivative of $\xi$ results in

$$\dot{\xi} = \alpha \mathbf{p}_0 \cdot \dot{\mathbf{r}} = \alpha p_0 v_0 = k v_0 = \Omega \tag{23}$$

where we have made use of the definition $k = \alpha p_0$. As expected, this parameter has the physical dimension of a wave vector and, if we introduce a wavelength $\lambda$ such that $\Omega = 2\pi v_0/\lambda$, then we find $k = 2\pi/\lambda$ and

$$p_0 = \frac{2\pi}{\alpha \lambda} \tag{24}$$

which coincides with Eq. (10) found before as a result of the periodic condition on the plane wave of Eq. (8). The result of Eq. (24) was clearly expected even though it stems from a classical reasoning, but its quantum nature is apparent as soon as the exact correspondence to the de Broglie relationship is invoked. It means that the parameter $\alpha$ equals the inverse of the reduced Planck constant $\hbar$ and it is easy to demonstrate that this equality plays a further role in the connection of Eq. (21) with the time-independent Schrödinger equation of a free particle of total energy $E = p_0^2/(2m)$. Indeed, by combining Eq. (21) and the definition of $\Omega$, it is found that

$$-\frac{1}{2m\alpha^2} \nabla^2 f = E f \tag{25}$$

and, if we set $\alpha = 1/\hbar$, we get the fundamental equation of wave mechanics for the free particle.

Having clarified the role of the second Newton's law in the derivation of the time-independent Schrödinger equation of a non-interacting particle, we are now ready to undertake the more difficult task of laying the foundation of wave mechanics by using the tool of Newtonian dynamics in presence of interaction.

## IV. CONSERVATIVE FORCES: FROM NEWTON TO SCHRÖDINGER

Most of the conclusions reached so far pertain to a region of space where the net force acting on the particle is negligible. Now, we abandon this strong simplification and, in this Section, we make the determined effort to deal with the difficult problem of finding the Schrödinger equation for a particle whose acceleration is described by the second Newton's law. As recalled in the introductive Section, the idea of an intimate relationship between Newton's laws of dynamics and Schrödinger equation dates back to 1927 when Ehrenfest came out with the well-known theorem that bears his name.[30] The theorem states that, given a quantum state represented by the Schrödinger wave function, then the expectation value of the time derivative of the momentum operator is equal to the expectation value of the negative gradient of the potential energy function. In other words, the quantum-mechanical expectation values reproduce the structure of the second Newton's law for conservative forces. The current attempt aims, instead, at a complete role reversal between hypothesis and thesis of the Ehrenfest theorem. Indeed, given Newton's laws, we try to derive the Schrödinger equation. The underlying assumption is the de Broglie momentum that establishes the necessary premise for the Schrödinger theory as well as the current work.

To fulfill the plan, we need to prove that Eq. (19) suits our purposes when the second Newton's law reads

$$\ddot{\mathbf{r}} = -\frac{\nabla U}{m} \qquad (26)$$

where the potential energy $U = U(\mathbf{r})$ is a function of the spatial coordinates only. The combination of Eqs. (19) and (26) is open to further treatment after the multiplication by $f(\xi)$ and three-dimensional spatial integration

$$\int f(\xi)\left[-\frac{(\mathbf{p}_0 \cdot \dot{\mathbf{r}})^2}{p_0^2}\nabla^2 f(\xi) + \frac{d^2 f(\xi)}{dt^2}\right]d\mathbf{r} = -\frac{\alpha}{m}\int[\mathbf{p}_0 \cdot \nabla U(\mathbf{r})]f(\xi)\frac{df(\xi)}{d\xi}d\mathbf{r}. \qquad (27)$$

Next, we apply the divergence theorem (or Gauss theorem)[32] useful for the integration by parts. This version is obtained when the theorem is applied to the product between a scalar function $u$ and a vector field $\mathbf{w}$, then

$$\int_V (\mathbf{w}\cdot\nabla u)d\mathbf{r} + \int_V u(\nabla\cdot\mathbf{w})d\mathbf{r} = \int_S u(\mathbf{w}\cdot\mathbf{n})dS \qquad (28)$$

where $\mathbf{n}$ is a unit vector orthogonal to the surface $S$ that contains the volume $V$.

Eq. (28) allows us to rewrite the integral appearing on the right-hand side of Eq. (27) according to

$$\int_V (\mathbf{p}_0\cdot\nabla U)f(\xi)\frac{df}{d\xi}d\mathbf{r} = \int_S U(\mathbf{r})f(\xi)\frac{df}{d\xi}(\mathbf{p}_0\cdot\mathbf{n})dS - \int_V U(\mathbf{r})\left\{\nabla\cdot\left[\mathbf{p}_0 f(\xi)\frac{df}{d\xi}\right]\right\}d\mathbf{r} \qquad (29)$$

Now, we note that the surface integral is vanishing when we let the surface $S$ go to infinite,

$$\int_S U(\mathbf{r})f(\xi)\frac{df}{d\xi}(\mathbf{p}_0\cdot\mathbf{n})dS \to 0. \qquad (30)$$

This happens by virtue of the vanishing values of $f(\xi)$ and its first derivative when $|\mathbf{r}| \to \infty$. The assumption of rapidly converging functions to zero at large distances is clearly satisfied for the wave functions appearing in Schrödinger theory and we use the identical assumption in this context too. Then, Eq. (27) can be recast as follows

$$\int f(\xi)\left[-\frac{(\mathbf{p}_0 \cdot \dot{\mathbf{r}})^2}{p_0^2}\nabla^2 f + \frac{d^2 f}{dt^2}\right]d\mathbf{r} = \frac{\alpha^2}{m}\int U(\mathbf{r})p_0^2 \frac{d}{d\xi}\left[f(\xi)\frac{df}{d\xi}\right]d\mathbf{r} \qquad (31)$$

where we have used the identity

$$\nabla \cdot \left[\mathbf{p}_0 \, f(\xi)\frac{df}{d\xi}\right] = \alpha p_0^2 \frac{d}{d\xi}\left[f(\xi)\frac{df}{d\xi}\right]. \qquad (32)$$

The result of Eq. (31) is obvious if we let the two kernels be the same or

$$f(\xi)\left[-\frac{(\mathbf{p}_0 \cdot \dot{\mathbf{r}})^2}{p_0^2}\nabla^2 f + \frac{d^2 f}{dt^2}\right] = \frac{\alpha^2 p_0^2}{m}U(\mathbf{r})\frac{d}{d\xi}\left[f(\xi)\frac{df}{d\xi}\right] \qquad (33)$$

and, at this stage, further elaboration of Eq. (33) deserves more explaining. Indeed, additional handling of Eq. (33) has to be made in view of the condition established in Eq. (14). The condition is crucial to make sure that classical and Schrödinger pictures refer to the same constant of motion as indicated in Eq. (4). This means that we look for a solution $f(\xi)$ that conforms with the prescription of Eq. (14) where the reference to the selected times $t_j$ is now made explicit by means of the following notation

$$\left\{ f(\xi)\left[-\frac{(\mathbf{p}_0\cdot\dot{\mathbf{r}})^2}{p_0^2}\nabla^2 f+\frac{d^2 f}{dt^2}\right]\right\}\bigg|_{t=t_j} = \frac{\alpha^2 p_0^2}{m}\left\{U(\mathbf{r})\frac{d}{d\xi}\left[f(\xi)\frac{df}{d\xi}\right]\right\}\bigg|_{t=t_j}. \quad (34)$$

Under this time restriction, when $\dot{\mathbf{r}}(t_j)=\mathbf{p}_0/m=\mathbf{v}_0$, Eq. (34) can be modified as follows.

First, $d\xi$ appearing on the right-hand side is rearranged according to $d\xi = \alpha\mathbf{p}_0\cdot d\mathbf{r} = \alpha\mathbf{p}_0\cdot\mathbf{v}_0\, dt$. In this way, setting $\Omega = \alpha p_0^2/m$, we find

$$\left\{\frac{d}{d\xi}\left[f(\xi)\frac{df}{d\xi}\right]\right\}\bigg|_{t=t_j} = \frac{1}{\Omega^2}\left\{\left(\frac{df}{dt}\bigg|_{t=t_j}\right)^2 + \left[f(\xi)\frac{d^2 f}{dt^2}\right]\bigg|_{t=t_j}\right\} \quad (35)$$

and Eq. (34) becomes

$$f(\xi)\big|_{t=t_j}\left\{-\frac{(\mathbf{p}_0\cdot\mathbf{v}_0)^2}{p_0^2}\left[\nabla^2 f\right]_{t=t_j} + \frac{d^2 f}{dt^2}\bigg|_{t=t_j}\right\} =$$
$$\frac{\alpha^2 p_0^2}{m\Omega^2}U(\mathbf{r})\big|_{t=t_j}\left\{\left(\frac{df}{dt}\bigg|_{t=t_j}\right)^2 + \left[f(\xi)\frac{d^2 f}{dt^2}\right]\bigg|_{t=t_j}\right\} \quad (36)$$

The second effect of the constraint $t=t_j$ is on the time derivatives. These can be calculated if we point out that around $t=t_j$ the momentum appears to obey the conservation law typical of the free particle mentioned at the end of the previous Section. In other terms, the function $f(\xi)$ has to guarantee that the momentum returns to the chosen value $\mathbf{p}_0$ at any $t_j$ so that $f(\xi_j) = f(\alpha\mathbf{r}_j\cdot\mathbf{p}_0) = Exp(i\alpha\mathbf{r}_j\cdot\mathbf{p}_0)$. The condition implies a restriction on the time dependence of the wave function $f(\xi)$ that should evolve according to a harmonically varying function between one $t_j$ and the other. This means that the time derivatives can now be calculated

$$\left.\frac{df(\xi)}{dt}\right|_{t=t_j} = i\Omega f(\xi)\big|_{t=t_j} \tag{37}$$

$$\left.\frac{d^2 f(\xi)}{dt^2}\right|_{t=t_j} = -\Omega^2 f(\xi)\big|_{t=t_j} \tag{38}$$

where the angular frequency $\Omega$ appears because of the time derivative of the variable $\xi$ at $t = t_j$. In the end, Eq. (36) can be transformed into

$$\frac{p_0^2}{m^2}\left[\nabla^2 f(\xi)\right]_{t=t_j} + \Omega^2 f(\xi)\big|_{t=t_j} = 2\frac{\alpha^2 p_0^2}{m}\left[U(\mathbf{r})f(\xi)\right]_{t=t_j} \tag{39}$$

and recalling that $\Omega = \alpha p_0^2 / m$, we obtain

$$-\frac{1}{2\alpha^2 m}\left[\nabla^2 f(\xi)\right]_{t=t_j} + \left[U(\mathbf{r})f(\xi)\right]_{t=t_j} = \frac{\Omega}{2\alpha} f(\xi)\big|_{t=t_j} \tag{40}$$

This equation is obeyed even if we look for solutions that are independent from the choice of the reference time $t_j$ and, by relaxing the time constraint, we end up with the following result

$$-\frac{1}{2\alpha^2 m}\nabla^2 f(\xi) + U(\mathbf{r})f(\xi) = \frac{\Omega}{2\alpha} f(\xi) \tag{41}$$

which is in close resemblance to the Schrödinger equation. The similarity is remarkable because Eq. (41) has been derived on the basis of classical arguments and shows many of the features that characterize the Schrödinger equation. What is more, the similarity becomes much clearer if we consider the de Broglie relationship that was used at the end of Section II to reach the conclusion about the correct value of $1/\hbar$ for the constant $\alpha$. In this instance, Eq. (41) is

$$-\frac{\hbar^2}{2m}\nabla^2 f(\xi) + U(\mathbf{r})f(\xi) = \frac{\hbar\Omega}{2}f(\xi) \qquad (42)$$

which seems to differ from the time-independent Schrödinger equation by only a factor of one-half on the right-hand side. This little difference will be brought into sharp focus in due course. For the time being, however, let us assume $\omega = \Omega/2$ and the final result is

$$-\frac{\hbar^2}{2m}\nabla^2 f(\xi) + U(\mathbf{r})f(\xi) = \hbar\omega f(\xi) \qquad (43)$$

Eq. (43) is the well-known time-independent Schrödinger equation and, to get back to the canonical time-dependent Schrödinger equation, we can now define a new wave function $\psi(\mathbf{r},t) = f(\xi)exp(-i\omega t)$ that incorporates the arbitrary function $f(\xi)$ we introduced at the very beginning. By doing so, we achieve the correct equation

$$-\frac{\hbar^2}{2m}\nabla^2 \psi(\mathbf{r},t) + U(\mathbf{r})\psi(\mathbf{r},t) = i\hbar\frac{\partial \psi(\mathbf{r},t)}{\partial t}. \qquad (44)$$

Nonetheless, the result would be incomplete if we were unable to explain why the factor of one-half makes its appearance on the right-hand side of Eq. (42) and, consequently, in the angular frequency $\omega = \Omega/2$ of Eq. (43). The explanation goes back to the crucial result of Eq. (40) that is relative to all the points of the trajectory where the momentum is recurrently $\mathbf{p}_0$. In manipulating that equation, we took advantage of $\Omega = \alpha p_0^2/m$ and now, having set $\alpha = 1/\hbar$, it results that $\Omega = p_0^2/(\hbar m)$. But, the particle energy $E$ for which Eq. (40) holds is the kinetic energy $E = p_0^2/(2m)$ relative to the times $t_j$ that characterize Eq. (14). By substitution, we find $\Omega = 2E/\hbar$. But, in the passage from Eq. (40) to Eq. (42), we have relaxed the time constraint and the particle energy must be generally defined by the Planck-Einstein relationship $E = \hbar\omega$ so that we obtain $\Omega = 2\omega$. In conclusion, the correct angular frequency expected in the time-independent Schrödinger equation is $\omega = \Omega/2$.

## V. CONCLUSIONS

To sum up, the Schrödinger picture of quantum mechanics is obtained from Newtonian dynamics on the basis of the existence of matter waves dictated by de Broglie relationship. The objective has been accomplished thanks to the correspondence between classical density functions and the tool of linear operators. The correspondence is such that the classical density functions can be represented by the squared modulus of eigenfunctions that solve a corresponding eigenvalue problem established for a given linear operator whose specific definition is unimportant for the argument developed here. Furthermore, the correspondence provides a fundamental condition for the representation of the eigenfunctions of a particle with a given translational momentum. More importantly, the application of the second Newton's law to the dynamical variables appearing in the eigenfunctions generates a wave equation. This one is introduced in the divergence theorem of vector calculus and, after some algebraic manipulation, the result is then constrained to satisfy the fundamental condition for the representation of the eigenfunctions of a particle with a given translational momentum. The time restriction is finally released and the Schrödinger equation appears in dependence of a parameter with physical dimensions of the inverse of an action. This free parameter has the same role played by the constant $K$ introduced by Schrödinger in his successful effort to calculate the energy levels of the hydrogen atom. Tuning the parameter on the correct value determined by the de Broglie condition on the momenta, the final result is exactly the Schrödinger equation.

---

*michele.marrocco@enea.it

[1] D. F. Styer *et al.*, ''Nine formulations of quantum mechanics,'' Am. J. Phys. **70**, 288–297 (2002).

[2] W. P. Schleich, D. M. Greenberger, D. H. Kobe, and M. O. Scully, "Schrödinger equation revisited," Proc. Natl. Acad. Sci. **110**, 5374-5379 (2013).

[3] E. Schrödinger, ''Quantisierung als Eigenwertproblem (Erste Mitteilung),'' Ann. Phys. **384**, 361–376 (1926); ''Quantisierung als Eigenwertproblem (Zweite Mitteilung),'' Ann. Phys. **384**, 489–527 (1926). Reprinted and translated in E. Schrödinger, *Collected Papers on Wave Mechanics* (Chelsea Publishing Company, New York, 1982).

[4] R. P. Feynman, R. B. Leighton, and M. Sands, *The Feynman Lectures on Physics* (Addison-Wesley, Reading, MA, 1977).

[5] R. P. Feynman, "Space-time approach to non-relativistic quantum mechanics," Rev. Mod. Phys. **20**, 367-387 (1948).

[6] D. Derbes, "Feynman's derivation of the Schrödinger equation," Am. J. Phys. **64**, 881-884 (1996).

[7] E. Nelson, "Derivation of the Schrödinger equation from Newtonian mechanics," Phys. Rev. **150**, 1079–1085 (1966).

[8] E. Santamato, "Geometric derivation of the Schrödinger equation from classical mechanics in curved Weyl spaces," Phys. Rev. D **29**, 216-222 (1984).

[9] B. Roy Frieden, "Fisher information as the basis for the Schrödinger wave equation," Am. J. Phys. **57**, 1004-1008 (1989).

[10] C. G. Gray, G. Karl, and V. A. Novikov, "From Maupertius to Schrödinger. Quantization of classical variational principles", Am. J. Phys. **67**, 959-961 (1999).

[11] W. E Lamb, "Superclassical quantum mechanics: The best interpretation of nonrelativistic quantum mechanics," Am J Phys **69**, 413–422 (2001).

[12] M. J. W. Hall and M. Reginatto, "Schrödinger equation from an exact uncertainty principle," J. Phys. A: Math. Gen. **35**, 3289–3303 (2002).

[13] L. Fritsche and M. Haugk, "A new look at the derivation of the Schrödinger equation from Newtonian mechanics," Ann. Phys. **12**, 371-402 (2003).

[14] G. Grössing, "From classical Hamiltonian flow to quantum theory: derivation of the Schrödinger equation," Found. Phys. Lett. **17**, 343-362 (2004).

[15] J. S Briggs, S. Boonchui, and S. Khemmani, "The derivation of time-dependent Schrödinger equations," J. Phys. A: Math. Theor. **40**, 1289–1302 (2007).

[16] G. Grössing, "The vacuum fluctuation theorem: Exact Schrödinger equation via nonequilibrium thermodynamics," Phys. Lett. A **372**, 4556-4563 (2008).

[17] N. Cufaro Petroni and M. Pusterla, "Lévy processes and Schrödinger equations," Physica A **338**, 824–836 (2009).

[18]P. R. Sarma, "Direct derivation of Schrödinger equation from Hamilton-Jacobi equation using uncertainty principle," Rom. J. Phys. **56**, 1053–1056 (2011).

[19]A. Deriglazov and B. F. Rizzuti, "Reparametrization-invariant formulation of classical mechanics and the Schrödinger equation," Am J Phys **79**, 882–885 (2011).

[20]J. H. Field, "Derivation of the Schrödinger equation from the Hamilton–Jacobi equation in Feynman's path integral formulation of quantum mechanics," Eur. J. Phys. **32**, 63–87 (2011).

[21]M. A. de Gosson and B. J. Hiley, "Imprints of the quantum world in classical mechanics," Found. Phys. **41**, 1415–1436 (2011).

[22]A. Caticha, "Entropic dynamics, time and quantum theory," J. Phys. A: Math. Theor. **44**, 225303 (2011).

[23]G. González, "Relation between Poisson and Schrödinger equations," Am J Phys **82**, 715–719 (2012).

[24]C. Cohen-Tannoudji, B. Diu, and F. Laloe, *Quantum Mechanics* (Wiley, New York, 1991).

[25]D. J. Griffiths, *Introduction to Quantum Mechanics* (Pearson Prentice Hall, Upper Saddle River, NJ, 2005).

[26]L. Susskind and A. Friedman, *Quantum Mechanics. The Theoretical Minimum* (Basic Books, New York, 2014).

[27]L. D. Landau and E. M. Lifshitz, *Quantum Mechanics: Non-Relativistic Theory* (Pergamon, New York, 1977).

[28]R. M. Eisberg and R. Resnick, *Quantum Physics of Atoms, Molecules, Solids, Nuclei, and Particles* (John Wiley & Sons, New York, 1985).

[29]A. Messiah, *Quantum Mechanics* (Dover Publications, New York, 1999).

[30]P. Ehrenfest, "Bemerkung über die angenäherte Gültigkeit der klassischen Mechanik innerhalb der Quantenmechanik," Z. Physik **45**, 455–457 (1927).

[31]H. Goldstein, C. Poole, and J. Safko, *Classical Mechanics,* (Addison-Wesley, San Francisco, 2001).

[32]P. M. Morse and H. Feshbach, *Methods of Theoretical Physics,* (McGraw-Hill, New York, 1953).